\documentclass[12pt,aasms4]{aastex}

\def\lsim{\lower.5ex\hbox{$\; \buildrel < \over \sim \;$}}
\def\gsim{\lower.5ex\hbox{$\; \buildrel > \over \sim \;$}}
\usepackage{graphicx}

\def\jnl@aj{AJ}
\ifx\revtex@jnl\jnl@aj\let\tablebreak=&&&\nl\fi
\received{25 February 2004}
\accepted{23 June 2004}

\slugcomment{To Be Published in {\em Astrophysical Journal Letters,} {\bf 611,} 10 August 2004}

\begin{document}

\title{THE GALACTIC $^{26}$Al PROBLEM \& THE CLOSE BINARY SNIb/c SOLUTION? }

\author{J. C. Higdon}
\affil{W. M. Keck Science Center, Claremont Colleges,
Claremont, CA 91711-5916; jimh@lobach.JSD.claremont.edu}

\author{R. E. Lingenfelter and R. E. Rothschild}
\affil{Center for Astrophysics and Space Sciences, University of
California San Diego, La Jolla, CA 92093; rlingenfelter@ucsd.edu
and rrothschild@ucsd.edu}

\begin{abstract}
The origin of the long-lived (1.07 Myr mean life) radioactive
$^{26}$Al, which has been observed in the Galactic interstellar
medium from its 1.809 MeV decay gamma-ray line emission, has
been a persistent problem for over twenty years. Wolf-Rayet (WR)
winds were thought to be the most promising source, but their
calculated $^{26}$Al yields are not consistent with recent analyses
of the 1.809 MeV emission from the nearest WR star and nearby OB
associations. The expected $^{26}$Al yield from the WR star exceeds by as
much as a factor of 3, that set by the 2-$\sigma$ upper limit on
the 1.809 MeV emission, while the WR yields in the OB associations
are only about 1/3 of that required by the 1.809 MeV emission.
We suggest that a solution to these problems may lie in $^{26}$Al
from a previously ignored source: explosive nucleosynthesis in
the core collapse SNIb/c supernovae of WR stars that have lost
most of their mass to close binary companions. Recent nucleosynthetic
calculations of SNIb/c suggest that their $^{26}$Al yields depend
very strongly on the final, pre-supernova mass of the WR star,
and that those with final masses around 6 to 8 M$_{\odot}$ are
expected to produce as much as 10$^{-2}$ M$_{\odot}$ of $^{26}$Al
per supernova. Such binary SNIb/c make up only a small fraction of
the current SNIb/c and only about 1\% of all Galactic core collapse
supernovae. But they appear to be such prolific sources that the
bulk of the present $^{26}$Al in the Galaxy may come from just a
few hundred close binary SNIb/c and the intense 1.809 MeV emission
from nearby OB associations may come from just one or two such
supernova. More extensive SNIb/c calculations of the $^{26}$Al
yields versus pre-supernova mass are clearly needed to test this
possibility.
\end{abstract}


\maketitle

\section{Introduction}

Observable diffuse Galactic 1.809 MeV line emission from the decay
of long-lived (1.07 Myr mean life) radioactive $^{26}$Al was
predicted (Arnett 1977; Ramaty and Lingenfelter 1977) from early
estimates (Schramm 1971) of the nucleosynthetic yields in explosive
carbon burning in core-collapse supernovae of about 10$^{-5}$
M$_{\odot}$/SN of $^{26}$Al. Assuming a Galactic supernova rate
of 1 SN every 30 yr, this yield suggested an average steady-state
radioactive mass of 0.3 M$_{\odot}$ of $^{26}$Al in the Galaxy.
The 1.809 MeV line emission was subsequently discovered by Mahoney
et al. (1982, 1983) with the high resolution gamma-ray spectrometer
on {\em HEAO 3} at an intensity of $\sim$ 5x10$^{-4}$ ph/cm$^2$ s str
from the inner Galaxy. This flux, confirmed by later observations,
is nearly an order of magnitude higher than that predicted, and
implies a steady-state Galactic mass of 3.1$\pm$0.9 M$_{\odot}$
of $^{26}$Al (e.g. Knodlseder 1999).

This much higher $^{26}$Al mass, together with a lack of
information about its spatial distribution and the uncertainties in
model predictions of $^{26}$Al yields, led to suggestions of a
variety of additional possible sources for $^{26}$Al, including the
winds of Wolf-Rayet (WR) stars, asymptotic giant branch (AGB) stars,
novae, and other transient sources. For a time, however, the yield
calculations of neon burning and neutrino interactions in SNII core
collapse of the massive ($>$ 25 M$_{\odot}$) stars without wind
losses appeared (e.g. Timmes et al. 1995) to be adequate to account
for the observed $^{26}$Al. But calculations (e.g. Schaller et al.
1992) of the evolution of these stars, showed that the SNII models
with hydrogen-rich envelopes were not appropriate in this mass range,
because their winds blow off their hydrogen envelopes leaving much
smaller WR stars, which are expected to end in SNIb/c instead. The
SNII yields (Timmes et al. 1995, Thielemann et al. 1996) of less
massive ($<$ 25 M$_{\odot}$) stars could account for no more than
about 1/8 of the observed $^{26}$Al, and calculations (Woosley,
Langer \& Weaver 1995) of the yields of the SNIb/c supernovae of
the small final mass stars that resulted from the expected large
WR wind losses in single stars, suggested that these stars were
also minor contributors. Thus, the deep dredging of the WR winds
themselves were explored as a possible major source. Early
calculations (Langer et al. 1995; Meynet et al. 1997), assuming
WR wind mass loss rates that were much larger than observations
now suggest, gave the yields could account for about 1/2 of the
observed $^{26}$Al, and very recent calculations (Vuissoz et al.
2004) using current wind loss estimates, but including the effects
of stellar rotation, now give even higher $^{26}$Al yields.

Studies of the spatial distribution of the Galactic $^{26}$Al
from COMPTEL, by Knodlseder et al. (1999ab) have also shown that
the diffuse 1.809 MeV line emission most closely correlates with
the distributions of young, massive stars. This clearly implies
that such stars are the source of the bulk of the Galactic $^{26}$Al,
and rules out novae, AGB stars and other older population sources.
This would also seem to support WR winds as the source, but other
recent observations argue against that.

\section{Problems with a WR Wind Source of $^{26}$Al}

First, the 1.809 MeV flux and $^{26}$Al yields that would be expected
from WR winds in the most recent calculations (Vuissoz et al. 2004)
exceed by as much as a factor of 3, the upper limits on the 1.809
MeV line from COMPTEL for the closest WR star, $\gamma^2$Velorum,
assuming (Oberlack et al. 2000; Pozzo et al. 2000) a distance of
258 to 410 pc. This star has an estimated (Schaerer et al. 1997)
initial mass of 57$\pm$15 M$_{\odot}$, and for a 60 M$_{\odot}$
WR star Vuissoz et al. (2004) calculate a $^{26}$Al wind yield of
2.24$\times$10$^{-4}$ M$_{\odot}$, with a maximum capture of 1/3
of that mass by its companion (e.g. Vanbeveren et al. 1998a).
Whereas the 2$\sigma$ upper limit of 1.1$\times$10$^{-5}$
photons/cm$^2$ s on the 1.809 MeV line flux from this star
(Oberlack et al. 2000) places a 2$\sigma$ upper limit of (0.6 to
1.5)$\times$10$^{-4}$ M$_{\odot}$, depending on the distance.

Second, recent analyses of the intense 1.809 MeV line fluxes
observed from the direction of the massive star formation regions,
Vela OB1, Cygnus OB2 and Orion OB1a, seem to further compound
the problem. Analyses of Vela OB1 by Lavraud et al. (2001) show
that, using the $^{26}$Al yields (Meynet et al. 1997) for large
WR wind losses, the expected 1.809 MeV emission from both
the WR winds and SNII was only about 1/5 of that observed. The
recent yields of Vuissoz et al. (2004) are only about 60\% higher
for the expected WR stars in this association, so the WR wind
yields are still only about 1/3 of that required. Similarly,
analyses of Cygnus OB2, using the Meynet et al. (1997) yields,
can account for only 1/2 of the observed emission (Knodlseder
et al. 2002; Pluschke et al. 2002), even after making a factor
of 3 increase over the number of observed O stars, as a correction
for obscuration. Although the recently calculated yields could
further close that gap, they already appear to be too high.
As we show below, the recent WR wind yields also fail to
account for the 1.809 MeV emission observed from Orion OB1a.

Lastly, the general problem is further complicated by the fact
that no 1.809 MeV emission, comparable to that from Vela OB1,
Cygnus OB2, or Orion OB1a, has been observed (Knodlseder et
al. 1999c) from the half dozen other equally large nearby OB
associations (Brown et al. 1996).

\section{The Solution: Close Binary SNIb/c?}

We suggest that the solution to the all of these $^{26}$Al
problems may lie in the new nucleosynthetic calculations of
SNIb/c by Nakamura et al. (2001) for larger final mass WR stars,
that are expected (Van Bever \& Vanbeveren 2003) to result from
mass transfer to close binary companions. Nakamura et al. (2001)
calculate for He cores without late mass loss that in final,
pre-supernova WR masses of 6 to 8 M$_{\odot}$ the $^{26}$Al
yields reach 6.7$\times$10$^{-3}$ to 1.2$\times$10$^{-2}$
M$_{\odot}$, while at high masses of 10 to 16 M$_{\odot}$ the
yields drop precipitously. Although the 6 to 8 M$_{\odot}$
yields might seem surprisingly large, such yields do seem to
be quite consistent with the very steep dependence of the
$^{26}$Al yield on final mass that Woosley et al, (1995)
found for SNIb/c of much smaller final masses expected
from the earlier large wind losses. They calculated SNIb/c
yields of 4.9$\times$10$^{-6}$ to 8.4$\times$10$^{-5}$
M$_{\odot}$ of $^{26}$Al for final masses ranging from 2.3
to 3.5 M$_{\odot}$, respectively, which can be approximated
by a power law in final mass to roughly the 6.5 power. Such
a power-law dependence would give a yield of 7.6$\times$10$^{-3}$
M$_{\odot}$ of $^{26}$Al at 7 M$_{\odot}$, which is quite
comparable to the SNIb/c values calculated by Nakamura et al.
(2001), as can be seen in Figure 1.

The relationship between the initial and final, pre-supernova
masses of WR stars is still uncertain. Perhaps the best determined
final masses are those for WR stars in close binaries, where the
mutual gravitational forces, Roche lobe overflow and common
envelope evolution are dominate (e.g. Paczynski 1971; Vanbeveren,
De Loore \& Van Rensbergen 1998a; Taam \& Sandquist 2000), and the
distribution of final masses depends most strongly on the range of
orbital parameters and stellar mass ratios, which have been extensively
measured (e.g. Popova, Tutukov \& Yungelson 1982; Duquennoy \& Mayor
1991). The final masses of single WR stars, however, depend (e.g.
Vanbeveren et al. 1998a; Maeder \& Meynet 2000) solely on the
radiation-driven wind loss rates, which vary strongly with the
changing stellar luminosity, mass, rotation and metallicity.

Currently the principal source of WR stars with final masses in the
peak $^{26}$Al producing range from 6 to 8 M$_{\odot}$ appears to be
those produced by mass transfer from massive stars in close binary
systems with orbital periods of 1 day to 10 yrs calculated by Van Bever
\& Vanbeveren (2003). Their calculated final masses, averaged over the
measured ranges of orbital period, angular momentum and stellar mass
ratios, are shown as a function of initial mass in Figure 2b. Such
binary systems appear to make up about 30\% of all binaries (e.g.
Duquennoy \& Mayor 1991), or 1/8 of all stars, assuming 0.4$\pm$0.1
(e.g. Popova et al. 1982) of all stars are in binaries.

Recent calculations by Meynet \& Maeder (2003) of the evolution of
single WR stars including rotation effects give final masses of $>$
10 M$_{\odot}$ (Fig.~2(b)). The SNIb/c explosions of such stars are
expected to have low $^{26}$Al yields which would also make them
minor contributors. Other recent calculations by Vanbeveren et al.
(1998b), however, suggest that a significant fraction of these stars
may also have lower final masses. Clearly further work is needed to
resolve this question and the 1.809 MeV observations will provide
important constraints.

Combining the $^{26}$Al yields (Woosley et al. 1995; Nakamura et al.
2001) for SNIb/c models as a function of pre-supernova mass of the
WR stars and the close binary mass transfer calculations (Van Bever
\& Vanbeveren 2003) of the pre-supernova mass versus the initial mass
of the WR stars in close binaries (Figs. 1 and 2b), we estimate the
expected $^{26}$Al yields for SNIb/c versus initial stellar mass.
These close binary SNIb/c yields are shown in Figure 2a, together
with estimates of the  yields from SNII, WR winds and single SNIb/c,
which also include those of longer period binaries. For the SNII
we take the yields recently calculated by Rauscher et al. (2002),
that include the new lower wind losses, for initial stellar masses
from 15 to 25 M$_{\odot}$, and thoes by Timmes et al. (1995) for
lower mass stars, where wind losses are not thought to be important.
For the WR wind yields we use those calculated by (Vuissoz et al.
2004). For the single SNIb/c and the longer period ($>$ 10 yr)
binaries, which should evolve like the single stars, we again take
the yields as a function of final WR mass (Fig. 1), together with
the Meynet \& Maeder (2003) calculations of that mass versus initial
mass with rotation effects (Fig. 2b) as a nominal value.

As can be seen (Fig. 2a), the dominant $^{26}$Al yield is expected
from SNIb/c of 30 to 50 M$_{\odot}$ WR stars in close binaries.
Stars in that mass range make up only about 8\% of all core-collapse
SN progenitors ($>$8 M$_{\odot}$), assuming a Salpeter (1955) initial
mass function (IMF), $dN/dM \approx M^{-2.35}$. Thus, those in close
binaries should make up only $\sim$ 1\% of all Galactic core-collapse
SN progenitors, since close binaries include only 1/8 of all stars.

\section{$^{26}$Al from Close Binary SNIb/c in the Galaxy}

We calculate the average steady-state mass of $^{26}$Al in the
Galaxy over its radioactive mean life of 1.07 Myr, by integrating
the $^{26}$Al yields (Fig. 2a) over a Salpeter IMF, and a total
core-collapse SN rate of 1 SN every 40 yr. We find that the close
binary SNIb/c contribute an average steady-state $^{26}$Al mass
of 2.5 M$_{\odot}$, while the WR winds, SNII and single SNIb/c
contribute 1.4, 0.4 and 0.14 M$_{\odot}$, respectively. For the
WR wind contribution, we also included the Galactic metallicity
enhancement, integrating the Galactic SN progenitor distribution
weighted by the metallicity squared, which is proportional to the
calculated WR wind yield. These contributions combine to give a total
steady-state mass of 4.4 M$_{\odot}$ of $^{26}$Al in the Galaxy,
which is somewhat higher than the 3.1$\pm$0.9 M$_{\odot}$ inferred
(e.g. Knodlseder 1999) from the Galactic 1.809 MeV emission.
However, as discussed above, the 1.809 MeV upper limit from the
nearest WR star suggest that the calculated WR wind contribution
may be too high by a factor of 2 or 3, which would also greatly
improve the agreement. Thus, more than 60\% of the $^{26}$Al is
expected to come from the SNIb/c of the 30 to 50 M$_{\odot}$
WR stars in close binaries. Since they make up only about 1\%
of the core-collapse SN progenitors, this contribution should
come from only about 300 SN out of the roughly 26,000 SN that have
occurred in the Galaxy over the last 1.07 Myr mean life of $^{26}$Al.

\section{$^{26}$Al from Close Binary SNIb/c in
Nearby OB Associations}

The expected average time dependent 1.809 MeV line emission from a
single OB association is shown in Figure 3 for a nominal moderate
sized OB association that initially contained 100 core-collapse SN
progenitors (8 to 120 M$_{\odot}$), determined from Monte Carlo
simulations using the mass dependent $^{26}$Al yields (Fig. 2a) and
the calculated (Schaller et al. 1992) stellar ages as a function of
initial mass. Because high-yield close-binary SNIb/c make up only
$\sim$ 1\% of the SN the probability of one occurring must also
be determined from Monte Carlo simulations for each individual OB
association, based on their size and age. Here we consider the
three nearby associations from the direction of which 1.809 MeV
line emission has been measured, Vela OB1, Cygnus OB2 and Orion OB1a.

Vela OB1 is a young, massive OB association at a distance of
1.8$\pm$0.4 kpc and an estimated age of 5$\pm$2 Myr (Lavraud et al.
2001), which is expected to have total of 118 to 155 core-collapse
SN progenitors for this age range, based on a Salpeter IMF and its
present population of 38 massive stars between 15 and 40 M$_{\odot}$
(Oberlack 1997). 1.809 MeV line emission of (2.9$\pm$0.6)$\times$10$^{-5}$
photons/cm$^2$ s was observed (Oberlack 1997) from that region with
COMPTEL. As can be seen (Fig. 4a), we find from Monte Carlo simulations
that such a flux would be expected from the $^{26}$Al in one or two
SNIb/c from close binary WR stars in that association about 40\%
of the time for ages between 5 and 6 Myr.

Cygnus OB2 is the largest of the OB associations in the Cygnus region
and the 1.809 MeV line emission of (5.8$\pm$1.5)$\times$10$^{-5}$
photons/cm$^2$ s observed from that direction by COMPTEL (Pluschke
et al. 2002; Knodlseder et al. 2002) is centered on it. Cygnus OB2
is at a distance of 1.7$\pm$0.4 kpc and its age is estimated to be
1-5 Myr (Comeron, Torra \& Gomez 1998; Herrero et al. 1999; Pluschke
et al. 2002; Knodlseder et al. 2002). We would expect at least 120
initial SN progenitors, based on optical observations of 40 O stars
(Massey et al. 1995). But since the region lies along a spiral arm,
it is highly obscured and the actual number may be significantly
larger (e.g. Knodlseder 2000). Nonetheless, from Monte Carlo
simulations of just the minimum number of progenitors (Fig. 4b),
we would expect 1.809 MeV line fluxes in the observed range from
the $^{26}$Al in one or two close binary SNIb/c in that association
about 20\% of the time if the age is in fact about 5 Myr, and that
probability would rise to about 50\% if even an additional 1/3 of
the O stars are unseen because of obscuration.

Orion OB1a is an older and smaller, but much closer association at a
distance of about 0.34 kpc and an age of 11.4$\pm$1.9 Myr with an
estimated initial 25 SN progenitors, based on the identification
(Brown, de Geus \& de Zeeuw 1994) of 53 stars between 4 and 15 M$_{\odot}$.
The 1.809 Mev flux observed (Diehl 2002 and personal communication
2004) by COMPTEL from this region was about (1 to 4)$\times$10$^{-5}$
photons/cm$^2$ s. From Monte Carlo simulations (Fig. 4c) of this
older association, we would expect contributions primarily from
the $^{26}$Al produced in SNII for ages $>$10 Myr which cannot
account for the observed flux, but for ages of 9.5 to 10 Myr we would
expect a residual contribution from an earlier close binary SNIb/c
about 30\% of the time which could account for the observed flux.

Lastly, we note that the stochastic nature of the emission also seems
to be supported by the fact that 1.809 MeV emission so far has been
found (Knodlseder et al. 1999c) from only 1/3 of the 9 (seen from
Vel OB1, Cyg OB2, and Ori OB1a, but not from Cep OB2, Gem OB1, Mon OB2,
Cen OB1, Ara OB1 and Sco OB1) largest, nearby ($<$2 kpc) OB associations
(Brown et al. 1996) in which $^{26}$Al from even a single high-yield
close binary SNIb/c could be seen. Such a fraction, however, is
consistent with the expectations of our Monte Carlo simulations.

Clearly more extensive calculations of the $^{26}$Al yields of SNIb/c
as a function of pre-supernova mass are needed to test such a source.

\acknowledgements

This work was supported by NASA's INTEGRAL Science Program via NASA grant
NAG5-12960.

\newpage


\begin{figure}
\includegraphics{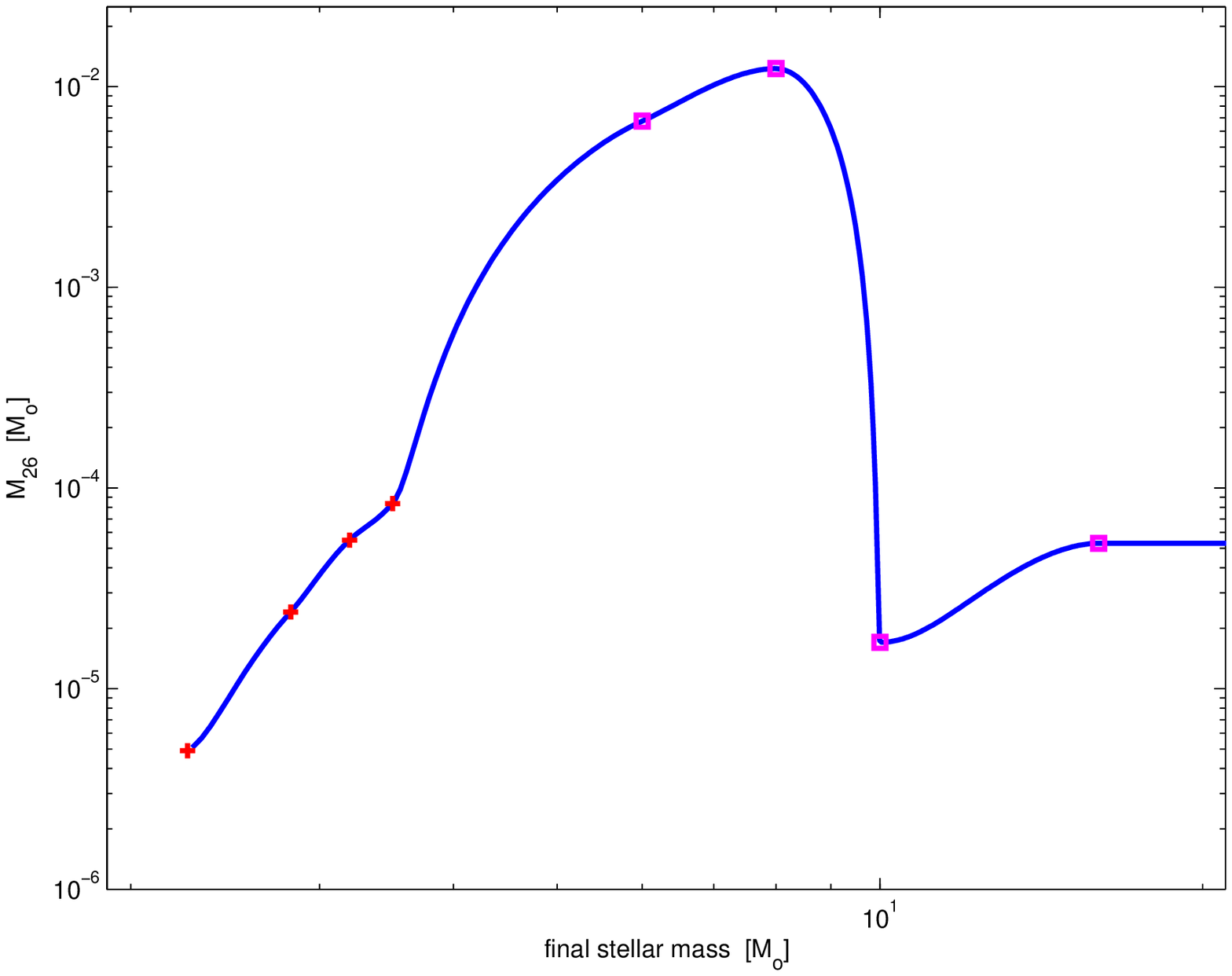}
\caption {The $^{26}$Al yield from SNIb/c explosions of WR stars
as a function of their final, pre-supernova mass, calculated by
Woosley et al. (1995) from 2.3 to 3.5 M$_{\odot}$ (crosses) and
Nakamura et al. (2001) from 6 to 16 M$_{\odot}$ (squares), showing
the high yields expected from the explosion of WR stars with final
masses around 6 to 8 M$_{\odot}$.}
\end{figure}

\begin{figure}
\includegraphics{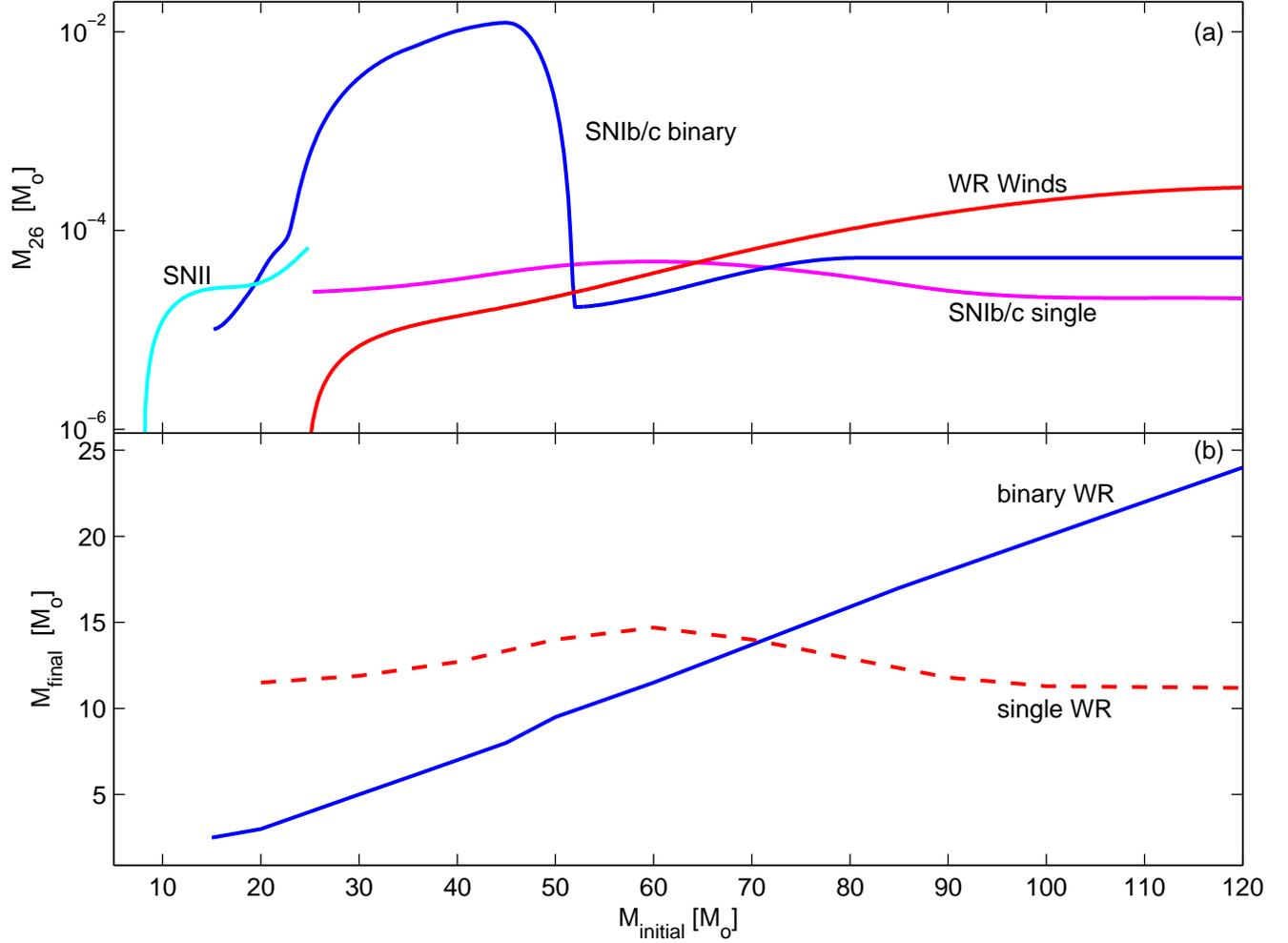}
\caption {(a) The $^{26}$Al yields from SNII, WR winds and SNIb/c
of both close binary and single WR stars as a function of their
initial, main sequence mass, showing the high yields expected
from close binary WR stars with initial masses around 30 to 50
M$_{\odot}$, and (b) the final, pre-supernova mass of WR stars
as a function of their initial, main sequence mass, calculated
for close binaries by Van Bever \& Vanbeveren (2003) and single
stars by Meynet \& Maeder (2003).}
\end{figure}

\begin{figure}
\includegraphics{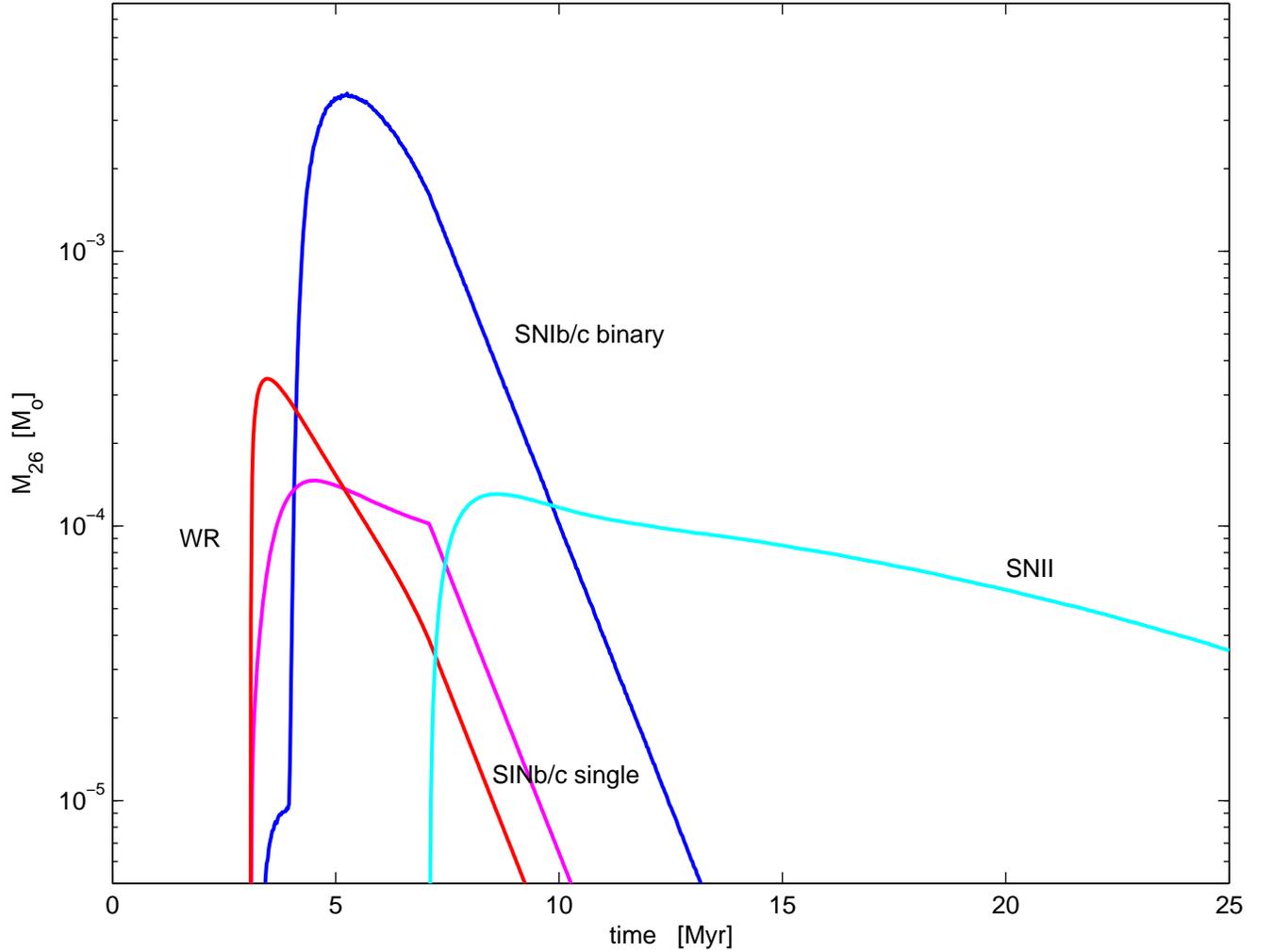}
\caption {The average accumulated mass of $^{26}$Al from SNII,
WR winds and SNIb/c of both close binary and single WR stars
in a nominal OB association starting with 100 SN progenitors
as a function of the age of the association. The actual time
dependence in a individual OB association is highly stochastic,
however, since the high yield, close binary SNIb/c make up only
$\sim$ 1\% of the core-collapse SN.}
\end{figure}

\begin{figure}
\includegraphics{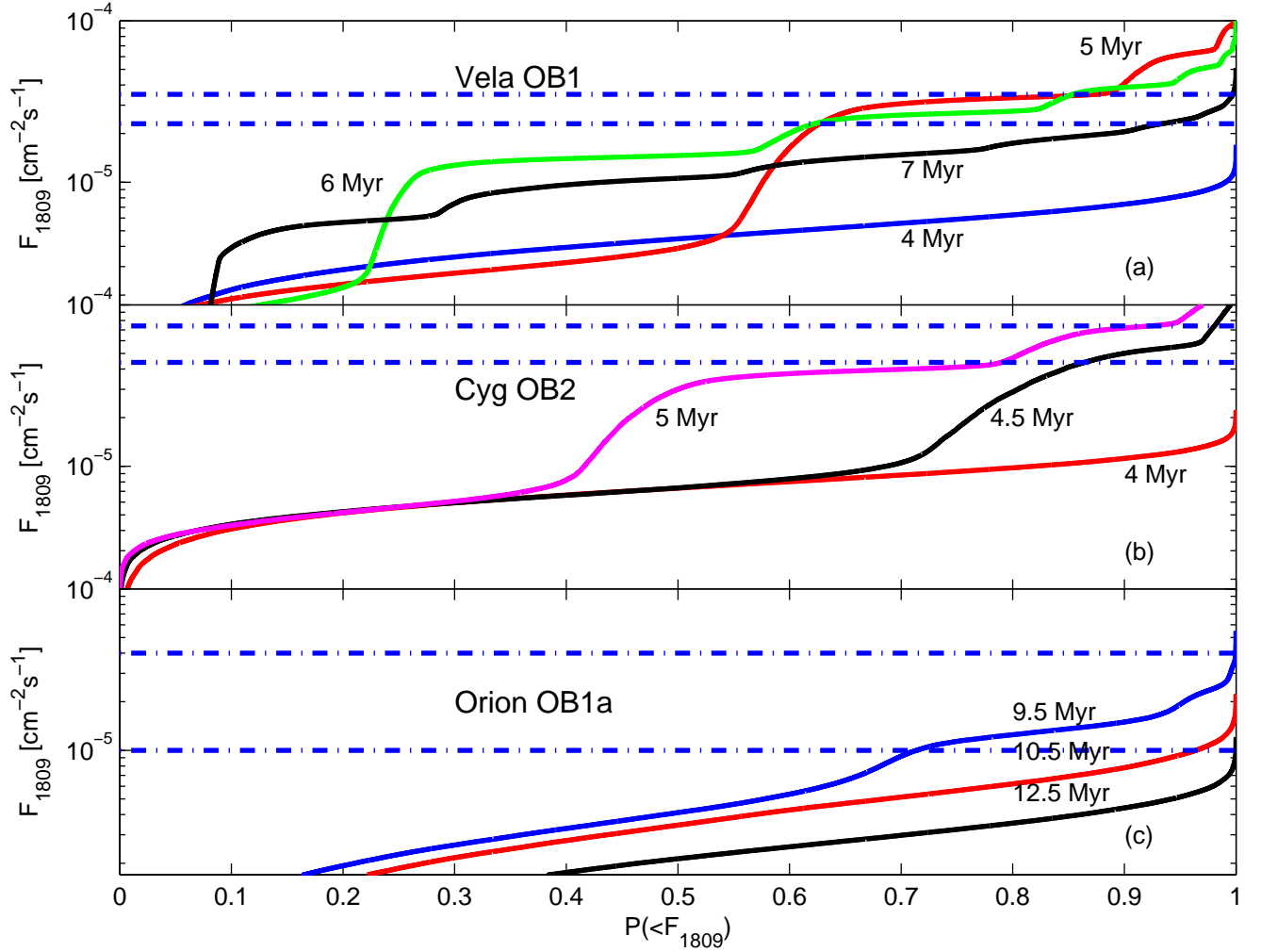}
\caption {Monte Carlo simulations of the probability distribution of
the 1.809 MeV line intensity from $^{26}$Al decay in the OB associations,
(a) Vela OB1, ages 4-7 Myr, (b) Cygnus OB2, ages 4-5 Myr, and (c) Orion
OB1, ages 9.5-12.5 Myr, for different assumed ages, showing a roughly
1/3 probability that just one or two close binary SNIb/c can account
for the observed flux from the directions of those associations.
Such a probability is also consistent with the fact that 1.809 MeV
emission has only been seen from 1/3 of the comparable OB associations.}
\end{figure}

\end{document}